\documentclass[twocolumn,prl,amsmath,amssymb]{revtex4}
\usepackage{graphicx}
\usepackage{dcolumn} 
\usepackage{bm}

\def\sss{$1_{\mbox{sss}}$}
\def\fss{$1_{\mbox{fss}}$}
\def\fsf{$1_{\mbox{fsf}}$}
\def\ffs{$1_{\mbox{ffs}}$}
\def\fff{$1_{\mbox{fff}}$}

\begin{document}   

\title{Inflation rule for Gummelt coverings with decorated decagons
and its implication to quasi-unit-cell models}
\author{Hyeong-Chai Jeong}
\address{Department of Physics, Sejong University,  
  Kwangjin-ku, Seoul 143-747, Korea \\} 
%\keyword{keyword}
%\PDBref[optional name]{refcode}
%\NDBref[optional name]{refcode}

\begin{abstract}
The equivalence between quasi-unit-cell models and Penrose-tile
models on the level of decorations is proved using inflation 
rules for Gummelt coverings with decorated decagons. 
Due to overlaps, Gummelt arrangement of decorated decagons gives 
rise to nine different (context-dependent) decagon decorations in 
the covering. The inflation rules for decagons for each of nine 
types are presented and shown that inflations from differently 
typed decagons always produce different decorations of inflated 
decagons. However, if the original decagon region is divided 
into ``equivalent'' rhombus Penrose tiles, typed-decagon 
arrangements in the tiles (of the same shape) become identical for 
the fourfold inflated decagons. This implies that a decagonal 
quasi-unit-cell model can be reinterpreted as a Penrose-tile model 
with fourfold deflated super-tiles. 
\end{abstract}

\maketitle                        % DO NOT DELETE THIS LINE

\section{Introduction}
	Quasicrystals are solids exhibiting long-range translational
order with a rotational symmetry that is forbidden in 
a periodic system ~\cite{Levine84}.
They have a quasiperiodic translational order and their lattice 
structures are called quasicrystalline lattices or quasilattices. 
There were several different ways of obtaining a quasilattice. 
Projection, dual grid, inflation, Penrose matching-rule and 
Gummelt overlapping-rule methods are some of known examples of 
obtaining a quasilattice~\cite{Janot92B,Gummelt96}.

The well-known two-dimensional (2D) Penrose lattice can be 
obtained by each of above methods. All of them make the equivalent
Penrose lattice structure but the basic building blocks 
they use to produce the structure are not the same. 
For example, the 5D hypercube, the two types of rhombic tiles and 
the decagon are the basic building blocks in the projection, arrow 
matching-rule and Gummelt overlapping-rule methods respectively. 
Therefore, the different ways of obtaining (the same) quasilattice 
structures
may produce different sets of atomic models, which are obtained 
by decorating each of basic building blocks identically and 
applying the operations to get the quasilattice structure. 
Therefore, equivalence in concept between two different 
approaches for quasicrystals should be investigated 
on two different levels, on the level of the lattice structures 
and on the level of atomic models.

Recently, we considered the relationship between 
rhombus-Penrose-tile (RPT) models, 
which are based on the arrow matching-rule method,
and decagonal quasi-unit-cell (dQUC) models,
which are based on the Gummelt 
overlapping-rule method~\cite{Jeong02JAP}.
An RPT model corresponds to decorating
each fat rhombus identically and each skinny rhombus
identically, and joining them to form a Penrose tiling.
Similarly, a dQUC model corresponds to 
decorating decagons identically and then covering
the plane according to the Gummelt overlapping rule. 
The decagon is called a quasi-unit cell (QUC) since 
its role corresponds to the role of unit cell in a periodic
crystal. It is similar to the unit cell in the sense that 
the QUC is a single repeating unit 
whose decoration determines the entire atomic structure of the solid. 
However, unlike the unit-cells, it overlaps its neighbors.  
Because of the overlaps, Gummelt arrangement of identically
decorated decagons can produce the context-dependent 
decoration of decagons in the covering. Therefore, the equivalence
between Penrose tiling and Gummelt Covering on the level
of lattice~\cite{Gummelt96,Steinhardt96,Jeong97} may not
guarantee the equivalence between RPT-models and dQUC models.

The lattice structure for real quasicrystals are often obtained by 
the 3D generalization of the planar Penrose lattice.
For example, the structure of decagonal quasicrystals can
be considered as a periodic stack of decagonal quasicrystalline 
planes. Therefore, the basic building blocks for real materials
should be 3D also. For decagonal quasicrystals, they are
two types of rhombic prisms for an RPT model~\cite{Jeong93}
and the decagonal prisms for a dQUC model~\cite{Steinhardt98Nat}.
Recently, we have constructed a dQUC model with the decagonal 
prism as the basic building block 
for $\mbox{Al}_{72}\mbox{Ni}_{20}\mbox{Co}_8$, 
one of the best-characterized 
quasicrystals~\cite{Steinhardt98Nat,Abe00}.
With the dQUC model, we could successfully reproduce 
the observed HRTEM images of 
$\mbox{Al}_{72}\mbox{Ni}_{20}\mbox{Co}_8$, 
as well as its measured stoichiometry, density and symmetry.
Very recent experiments on the surface images of the
above decagonal quasicrystals~\cite{McGrath02JAP}
are also explained 
more naturally with dQUC approaches 
than with conventional RPT models.
Furthermore, dQUC models provide simpler 
theoretical explanations for the existence of 
quasicrystals~\cite{Steinhardt96}. For RPT models,
two or more clusters analogous to Penrose tiles are needed 
and the complex atomic interactions are required to mimic the arrow
matching rules while the dQUC models use only a single 
type of basic building blocks as for crystals. 
Therefore, dQUC models seem to provide
physically more natural explanation for the quasicrystal 
formation and structures. 

The mathematical questions on the relationship between 
the set of dQUC models and the set of
conventional atomic models, such as RPT models
or hyper-cubic decoration models, are not fully investigated yet.
In our recent paper, we showed that an RPT model
is a dQUC model with the same edge size but
the converse is not true~\cite{Jeong02JAP}. 
Some dQUC models cannot be interpreted as
RPT models with the same edge length. 
However, we conjectured that a 
dQUC model can be an RPT model with
different edge size.
Here, we provide a mathematical proof for 
this claim. We first present the inflation rule for
a dQUC model and show that a dQUC 
model is an RPT model with the super-tiles whose
edge is $\tau^4$ times longer than the edge of
the quasi-unit decagon, where
$\tau = \frac{1+\sqrt{5}}{2}$. 

\section{Inflation rule for decorated decagons}

Real quasicrystals are 3D and therefore 
the basic building block for a dQUC model for real materials should 
be a decagonal prism. However, here we mod out the periodic direction
and take a 2D regular decagon as the basic building block. 
The 2D Gummelt covering then represents the lattice structure 
of each quasicrystalline plane
and the atomic decoration of the basic decagon is obtained
by the projection of atoms in the basic decagonal prism of
the model for a real quasicrystal.
For a classification of full 3D patterns from 
Gummelt prism decoration, see, Lord and 
Ranganathan~\cite{Lord01}.

%%=====(fig1: f_9type)=====
%%====== Figure 1 ============
\begin{figure}[h]
\includegraphics[width=80mm]{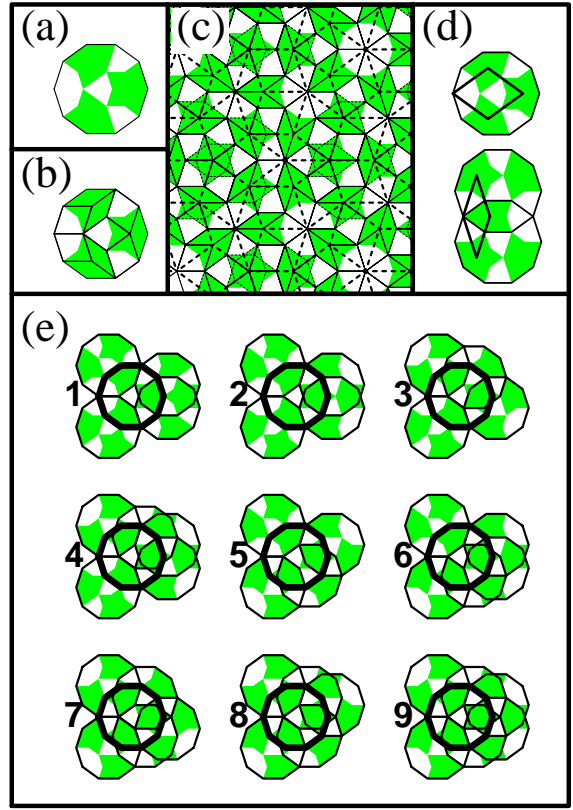} 
\caption{
(a) A marked regular decagon, the basic building block 
for a Gummelt covering. The decagon is marked with shaded
to represent the Gummelt overlapping rule. 
In the overlapped region between two neighboring 
decagons, shaded part from one decagon should
be also shaded from the other decagon.
(b)~Jack inscribed in a decagon. 
(c)~A Gummelt covering can be mapped into a Penrose tiling
by dividing up central area of decagon into Jack (thin solid lines).
A deflated Penrose tiling (thick dashed lines) is obtained
if we place a deflated skinny tile on every pair of
decagons and replace each decagon by a deflated fat tile 
as show in (d).
(e)~Nine ways of ``surrounding'' a decagon appeared
in a Gummelt covering.   
The centered decagon (thick lines) in the $k$-th ($k=1,\ldots,9$) 
surrounding configurations is called a type-$k$ decagon. 
}
\label{f_9type}
\end{figure}

In 1996, Gummelt showed that a 2D quasicrystalline structure with 
the decagonal symmetry can be obtained when the decagons,
shown in Fig.~\ref{f_9type}(a), are arranged with 
a specific overlapping rule~\cite{Gummelt96}. 
The overlapping rule demands 
that two decagons may overlap only if the shaded regions overlap. 
An infinite arrangement of decagons according to the Gummelt
overlapping rule is called a Gummelt covering. 
If we replace every decagon in the covering
by a Jack~\cite{Gardner77} as described in Fig.~\ref{f_9type}(b),
we obtain a Penrose tiling with rhombi whose edge length
is the same as that of the decagon as shown in 
solid lines in Fig.~\ref{f_9type}(c).
Due to the self-similar properties of the Gummelt covering,
we can obtain a Penrose tiling with other scale from the same 
covering.
For example, if we place a skinny tile on every
pair of decagons and replace each decagon by a fat tile 
as shown in Fig.~\ref{f_9type}(d),  
we get another Penrose tiling with ``deflated''(bigger) tiles
as shown in thick dashed lines in Fig.~\ref{f_9type}(c).

The mathematical equivalence between the lattice structures 
of the Penrose tiling and the Gummelt covering can be shown by considering 
the nearest-neighbor configurations
of decagons allowed by the overlapping rules~\cite{Jeong97}.
There are 20 different ways of surrounding a decagon with 
neighboring decagons, 
where surrounding a decagon means the edges of the decagon is covered 
by the interior of neighboring decagons.
These 20 configurations are equivalent to the 
configurations of ``surrounding'' a fat rhombus with
arrow edged rhombus tiles~\cite{Jeong97}. 
Among these 20 configurations of surrounding decagons, only 
9 configurations, shown in Fig.~\ref{f_9type}(e), appear in 
a  Gummelt covering.
The centered decagon in the $k$-th surrounding 
configurations of Fig.~\ref{f_9type}(e) is called a type-$k$ decagon 
where $k=1,\cdots,9$.

Recall that a dQUC model is defined by an atomic 
arrangement resulted from the covering of identically 
decorated decagons which are arranged according to 
the Gummelt overlapping rule. 
Although we arrange the identically decorated decagons,
differently-typed (or different context) decagons can be 
decorated differently in the covering since they have 
different overlaps with the neighboring decagons.

\begin{figure}[!h]
\includegraphics[width=80mm]{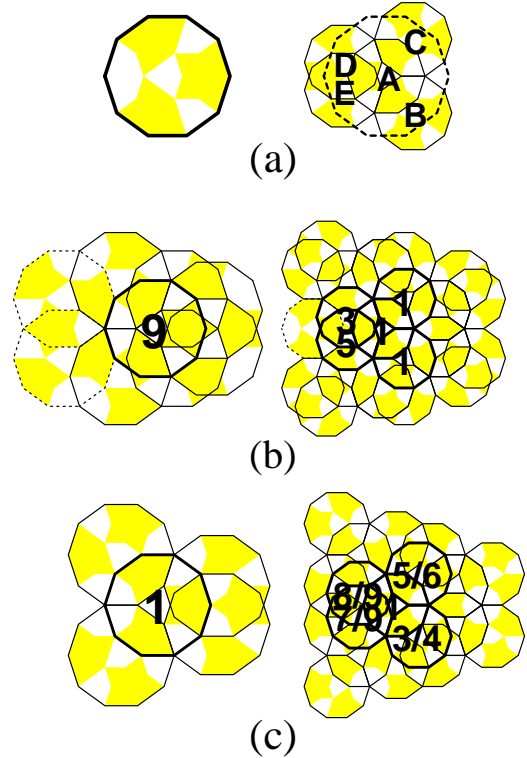} 
%\centerline{\psfig{file=f2.eps,width=80mm}}
\caption{
(a) Inflation of a decagon is defined as replacement
of a decagon (left) to the 5 smaller decagons shown at the right. 
The letters A,$\cdots$,E at the centers of the 
inflated decagons are shown for later references.
(b) Types of inflated decagons for a type-9 decagon.
Inflations of surrounding (thin solid lines) 
and accompanied (dotted lines) decagons determine 
the types of the inflated decagons.
The types of the inflated decagons at the position
A,B,C,D, and E are 1,1,1,3, and~5 respectively.
(c) Types of inflated decagons for a type-1 decagon.
In this case, there are no accompanied decagons.
Inflations of surrounding decagons are not enough for 
unique determination of the inflated decagon types. 
In addition to the position of 4 surrounding neighboring 
decagons, positions of 6 next-nearest-neighboring decagons
are needed to fix the types of inflated decagons.
See Fig.~3.
}
\label{f_inf}
\end{figure}

A Gummelt covering is self-similar as a Penrose tiling is
and therefore allows an inflation operation. 
The ``inflation of a decagon'' is a transformation
in which a decagon is replaced by 5 decagons as shown
in Fig.~\ref{f_inf}(a).
The edges of inflated decagons are $1/\tau$ times smaller 
than those of the original decagon. 
The letters A,$\cdots$,E 
represent the positions of the inflated decagons.
An infinite number of iterations of inflations 
(together with rescaling in the length by $\tau$)
produce a Gummelt covering. 
Inflation of a decagon covering means that the
inflation of every decagon in the entire covering.  
Inflation of a Gummelt covering produces another Gummelt covering. 

%%=====(Fig2-inflation)=====
%%====== Figure 2 ============

The types of inflated decagons are determined by the type of the 
original decagon. The inflated decagon at the position~A is 
always type-1 by the definition of the inflation.
The types of the inflated decagons at the other 4 positions
depend on the type of original decagon.
Unless the original decagon is type-1,
they are uniquely determined for a given type of the original decagon.
Note that a typed decagon implies a finite number of 
decagons at fixed neighboring positions. At least, the $k$-th surrounding 
configuration~(Fig.~\ref{f_9type}(e)) is always implied to the 
type-$k$ decagon as
the ``nearest neighbors''. In addition to this surrounding decagons,
``accompanied decagons'' as the ``next nearest neighbors'' are 
also implied for most cases (except type-1). 
``Accompanied decagons'' of the configuration-$k$ are
the decagons which are forced to be there in a  Gummelt 
covering for the given configuration-$k$.   
Figure~\ref{f_inf}(b) shows the case of the type-9 decagon. 
Six decagons drawn with thin solid lines are the surrounding 
decagons of the type-9 decagon. Two decagons (dotted-line) 
at the left of the figure are some of accompanied decagons.
The inflated decagons of the type-9 decagon 
(thick solid line decagons at the right figure of 
Fig.~\ref{f_inf}(b))
are fully surrounded by the inflated decagons of the
surrounding (thin solid lines)
and accompanied (dotted line) decagons of the original 
type-9 decagon. 
Inflations of surrounding 6 decagons force the inflated decagons 
at the positions B and C to be type-1. The inflated decagons at 
the positions D and E are not fully surrounded by 
the inflated decagons (solid lines) of the original surrounding decagons. 
However, inflations of the accompanied decagons of the configuration-9 
decagon give rise to the inflated decagon (dotted-line) at the left
and the D and E inflated decagons become type-3 and~5 respectively.
Similar mechanisms determine the types of inflated
decagons uniquely for the type-2,$\cdots$,type-9 decagons
and their types are given at table~1. 
(Subtypes of the type-1 decagons (\fss,$\cdots$,\fff) in 
the table are explained later.)

%%table1 here
%%=====(Fig3-sub_types)=====

%\begin{table}
%\caption{Types of Inflated decagons} 
\begin{tabular}{||c|c||c|c|c|c|c||}
\multicolumn{7}{c}{Table I.  Types of Inflated decagons} \\ \hline \hline
\multicolumn{2}{||c||}{Decagon}&
\multicolumn{5}{c||}{Inflated decagons} \\ \cline{3-7}
\multicolumn{2}{||c||}{Type}& A & B	& C	& \ D \	& E \\ \hline \hline
	&\sss		& \sss	& 3	& 5	& 8	& 7	\\ \cline{2-7}
	&\fss		& \sss	& 3	& 5	& 9	& 9	\\ \cline{2-7}
  1	&\fsf		& \sss	& 4	& 5	& 9	& 9	\\ \cline{2-7}
	&\ffs		& \sss	& 3	& 6	& 9	& 9	\\ \cline{2-7}
	&\fff		& \sss	& 4	& 6	& 9	& 9	\\ \hline
\multicolumn{2}{||c||}{2}& \fss	& 2	& 2	& 9	& 9	\\ \hline
\multicolumn{2}{||c||}{3}& \fsf	& 2	& \ffs	& 4	& 8	\\ \hline
\multicolumn{2}{||c||}{4}& \ffs	& 2	& \fff	& 4	& 8	\\ \hline
\multicolumn{2}{||c||}{5}& \ffs	& \fsf	& 2	& 7	& 6	\\ \hline
\multicolumn{2}{||c||}{6}& \ffs	& \fff	& 2	& 7	& 6	\\ \hline
\multicolumn{2}{||c||}{7}& \fff	& \fff	& \ffs	& 3	& 5	\\ \hline
\multicolumn{2}{||c||}{8}& \fff	& \fsf	& \fff	& 3	& 5	\\ \hline
\multicolumn{2}{||c||}{9}& \fff	& \fff	& \fff	& 3	& 5	\\ \hline
\hline
\multicolumn{7}{c}{}  
\end{tabular}
%\end{table}

\begin{figure}
\includegraphics[width=80mm]{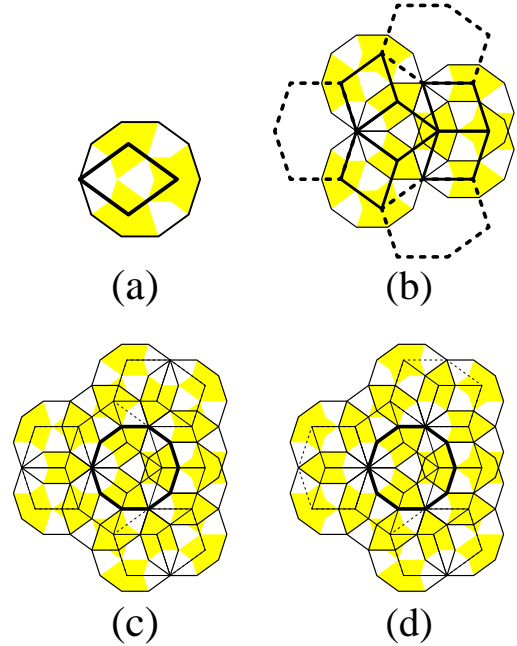} 
%\centerline{\psfig{file=f3.eps,width=80mm}}
\caption{
(a) Deflated fat rhombus inscribed in a decagon.
(b) C$'$-decagon consists of a Jack and three hexagons 
of deflated tiles.
A hexagon is composed of two fat rhombi and one skinny rhombus.
Depending on the orientation of the hexagons,
type-1 decagons are further classified by 8 
subtypes (\sss, $\ldots$, \fff).
(c) Subtype \sss \ decagon.
(d) Subtype \fsf \ decagon.
}
\label{f_subtype}
\end{figure}

This is not the case for the type-1 decagon. 
Unlike the other types, there are no accompanied decagons 
at the next nearest neighbors for the type-1 decagon. 
As shown in 
Fig.~\ref{f_inf}(c), there are 4 surrounding decagons in this case 
and their inflations are not enough for the unique determination of 
the inflated decagon types. 
Note that the configuration-1 in Fig.~\ref{f_9type}(e) 
corresponds to 
a Jack configuration with deflated tiles
if a deflated fat rhombus is 
inscribed in each decagon as in 
Fig.~\ref{f_subtype}(a)~\cite{Jeong97}. 
In a Penrose tiling, a Jack configuration 
is always a part of the C$'$-decagon~\cite{foot1,Jeong97}
consists of a Jack and three hexagons as shown in Fig.~\ref{f_subtype}(b).
There are two possible configurations for filling each
hexagon, $s$ and $f$ orientation. 
Each hexagon consists of three tiles, one skinny tile and two
fat tiles and the skinny tile in an $s$[$f$]-oriented hexagon
is at the side near [far from] the center of the C$'$-decagon. 
This gives rise to 8 subtypes for the type-1 decagon;
from $1_{\mbox{sss}}$ to $1_{\mbox{fff}}$.
The first, second, and third subscripts represent 
the orientations of the front, bottom, and top
hexagons respectively. 
Fig.~\ref{f_subtype}(c) and~(d) shows examples
of subtyped decagons, \sss \ (c) and \fsf \ (d).
All eight configurations satisfy the Gummelt overlapping rule
locally but the subtype $1_{\mbox{sfs}}$, $1_{\mbox{ssf}}$, 
and $1_{\mbox{sff}}$ decagons never appear in a Gummelt covering
since adding decagons to these configurations forces
a violation of the overlapping rule somewhere. 
The subtypes of the type-1 decagon fix
the positions of the 6 next-nearest-neighboring decagons 
(in the three hexagons)
in addition to the 4 surrounding decagons and their inflations
determine the types of the inflated decagons uniquely. 

%%====== Figure 3 ============

Table~1 shows the types of the inflated decagons for each type of 
the original decagon. The letters A,$\cdots$,E denote
the inflated decagons at the different positions 
denoted by Fig.~\ref{f_inf}(a).

\section{Relationship between dQUC models and RPT models}

\begin{figure}[!t]
\includegraphics[width=80mm]{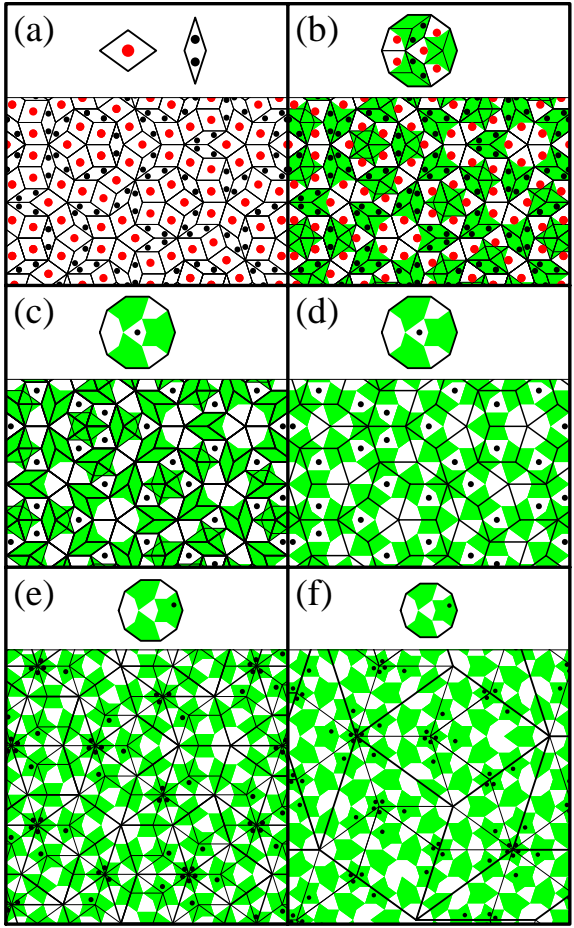} 
%\centerline{\psfig{file=f4.eps,width=80mm}}
\caption{
Relationship between dQUC models and RPT models.
An RPT atomic model is obtained by decorating
each fat rhombus by one atom and each skinny rhombus 
by two atoms as shown in the upper panel of (a) and 
joining them to form a Penrose tiling.
The RPT model of (a) is a dQUC model since the same atomic 
model can be obtained by decorating the decagon by 9 atoms 
as shown in the upper panel of (b) and arrange them to form 
the Gummelt covering. 
The dQUC model of (c) is not an RPT model with the same edge size
since some fat rhombi have atoms while the others don't.
However, it is an RPT model with the singly deflated Penrose 
tiles as illustrated in (d). For the dQUC model of (e),
the singly deflated fat rhombi (thin lines) are not 
decorated identically. Neither doubly (thick lines (e))
nor triply (thin lines in (f)) deflated rhombus tiles
are decorated identically.
However, it is an RPT model with fourfold deflated rhombus tiles 
as shown by thick lines (for clarity, only few super-tiles 
are shown).
}
\label{f_atom_deco}
\end{figure}

In this section, we compare the set of RPT models
with the set of dQUC models and show that two sets
are the same. It is quite easy to show that the 
set of dQUC models includes the set of RPT models.
We can easily construct an equivalent dQUC model 
for a given RPT model. Recall that a Gummelt covering 
is obtained when we replace each Jack in a Penrose tiling 
by a decagon as illustrated in Fig.~\ref{f_9type}(c).
For a given RPT model, by inscribing the ``Jack decoration''
(obtained by a Jack configuration of the decorated 
tiles of the RPT model) to the decagon, 
an equivalent dQUC model is obtained~\cite{Jeong02JAP}
as illustrated in Fig.~\ref{f_atom_deco}(a) and (b). 
The atomic arrangement resulted from the Penrose arrangement
of the decorated tiles in (a) is identical
to that results from the covering of the decorated decagons
in (b).

%%====== Figure 4 ============

However, it is not obvious if there is always an RPT model 
which is equivalent to a given dQUC model. Here, we show that 
it is the case by explicitly constructing such an RPT model 
using the inflation rule for decorated decagons.

Before we discuss the relationship between the set of all RPT models 
and that of all dQUC models, let us first consider restricted sets of models
whose basic building blocks (tiles or decagon) have the same edge length.
As mentioned before, the length of the decagon edges of the dQUC model,
obtained by the ``Jack decoration'' of an RPT model, is the same as the 
length of the tile edges. That is, for any RPT model, there is an equivalent 
dQUC model with the decagons whose edge length is the same as the 
tile edge length. However, the converse is not true.
Figure~\ref{f_atom_deco}(c) shows a counter example 
in which a decagon is decorated with an atom at the center. 
When the decagons in the coverings are resolved into a Jack,
some fat rhombi have atoms whereas some others do not.
Yet, it is an RPT model with the (singly) deflated Penrose tiles 
as illustrated in Fig.~\ref{f_atom_deco}(d). Each deflated fat 
rhombus is decorated 
identically (with an atom) and so is each deflated 
skinny rhombus (with no atom). 
However, this is rather accidental case for the decagon decoration 
of Fig.~\ref{f_atom_deco}(c). For the example of 
Fig.~\ref{f_atom_deco}(e), the singly deflated 
fat rhombi (thin lines) are
decorated with zero, one or two atoms depending on the context.
Neither doubly (thick lines in Fig.~\ref{f_atom_deco}(e))
nor triply (thin lines in Fig.~\ref{f_atom_deco}(f))
deflated rhombus tiles are decorated identically illustrated. 
For example, a doubly deflated fat tile has 1 or 2 atoms
and a triply deflated fat tile has 5 or 6 atoms depending
on the context. Therefore, in general, 
a dQUC model is not an RPT model with tiles up to triply deflated
tiles. However, the dQUC model of Fig.~\ref{f_atom_deco}(e) is an
RPT model with fourfold deflated rhombus tiles as shown in the
thick lines in Fig.~\ref{f_atom_deco}(f). The (fourfold-deflated) 
super-tiles of the same shapes are decorated identically. 
(for clarity, only few super-tiles are shown here
but we could not find any super-tiles decorated
differently even in much larger sample).
Now, the question is whether this is also accidental result
for the example of (e) or true for a general atomic decorations.

%%====== Figure 5 ============
\begin{figure}[!t]
\includegraphics[width=80mm]{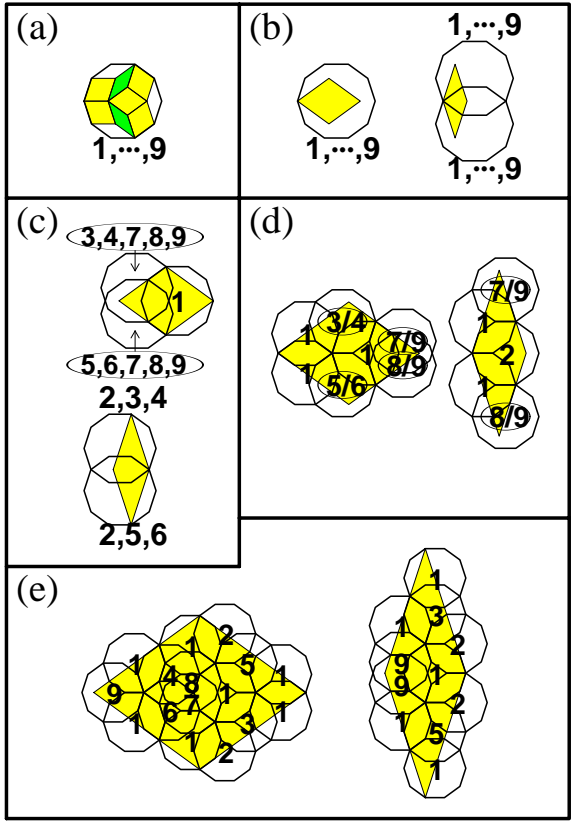} 
%\centerline{\psfig{file=f5.eps,width=80mm}}
\caption{
Deflated Penrose tiles and typed decagons.
The numbers at the center of or near 
the decagons represent their decagon types.
The positions and types of the decagons 
in the deflated tiles are determined by
the inflation rule of dQUC. 
(a)~The original rhombi and a decagon with the same edge size. 
The rhombi can be decorated differently depending 
on their position in the decagon.
(b)~Singly deflated rhombi and typed decagons.
The types of the decagons which cover the deflated 
rhombi are not uniquely determined. They can be any of 9 types.
(c)~Doubly deflated rhombi and typed decagons. 
The types of the decagons which cover the doubly deflated 
rhombi are not uniquely determined. For example,
the type of the decagon which covers the lower part
of the skinny shaped tile can be type-2, 5 or~6.
(d)~Triply deflated rhombi and typed decagons. 
The types of the decagons which cover the triply deflated 
rhombi are not uniquely determined. For example,
the type of the decagon which covers the lowest part
of the skinny shaped tile can be type-8 or~9.
(e)~Fourfold deflated rhombi and typed decagons.
The types of decagons are uniquely determined for both fat and 
skinny shapes.
}
\label{f_tile_deco}
\end{figure}

Below, we show that a dQUC model is an RPT model with 
the (fourfold-deflated) super-tiles in general. 
Note that the decagons with the same type are decorated 
identically even 
in the covering since their surrounding decagon arrangements
are the same. Therefore, deflated tiles are decorated identically 
if the types of the decagons at the same positions in all deflated 
tiles are the same. Therefore, we now consider the possible decagon 
types in the deflated tiles and see if they are unique.
Let us begin with singly deflated tiles. We know that a Penrose 
tiling with deflated tiles is obtained if we place deflated fat 
and skinny tiles
in decagons as shown in Fig.~\ref{f_9type}(d).
However, a general dQUC model cannot be an RPT model with 
(singly) deflated tiles since they are not decorated identically.
We already presented such an example in Fig.~\ref{f_atom_deco}(e).
In general, singly deflated fat tile are decorated differently
since the types of the decagons which cover the deflated tiles
are not uniquely determined as illustrated in 
Fig.~\ref{f_tile_deco}(b). A singly deflated fat tile 
(in the deflated tiling) is always covered by a decagon 
as shown in Fig.~\ref{f_tile_deco}(b)
but the decagon types can be any of nine types. 
Since the differently typed decagons are decorated 
differently for a general dQUC model, deflated fat tiles 
are decorated differently. Due to similar reasons,
deflated skinny tiles are decorated differently in general.
A singly deflated skinny tile is always covered by a pair of 
decagons by the way shown in Fig.~\ref{f_tile_deco}(b) but
their types can be any of nine. 

To investigate the relationship between multi-time deflated 
tiles and the typed decagons, we consider deflations of decagons. 
Deflation of decagons is the inverse process of 
inflation defined by Fig.~\ref{f_inf}(a).
A Gummelt covering is deflated if we substitute 
each configuration-1 in the covering by a 
deflated decagon whose edge length is $\tau$ times larger
than that of the original decagons.
If each deflated decagon in the deflated covering
is replaced by a Jack then a Penrose tiling 
with deflated tiles is obtained. 
If deflated decagons are replaced by doubly deflated tiles
as shown in Fig.~\ref{f_9type}(d) and
overlayed with the original decagons, the relationship
between doubly deflated tiles and typed decagons are 
obtained. 

Conversely, if we draw a Jack to the original decagon
and inflate a decagon, relationship
between the tiles (of the original size) and 
the inflated decagons are obtained. 
This relationship should be the same as the relationship between 
deflated tiles and decagons (of the original size)
due to the self-similarity of the covering. 
All possible decagon types 
in a deflated tile is then obtained by considering all 
thirteen different cases in the table. The relationship between 
multi-time deflated tiles and the typed decagons can be obtained 
by inflating decagons multi times.
Figures~\ref{f_tile_deco}(c)--(e) show the relationship
between (multi-time) deflated tiles and the 
typed decagons obtained this way. 
Decagon positions in the tiles of the 
same shape are always identical but their types 
are not uniquely determined for doubly ((c)) and 
triply ((d)) deflated tiles. For example, doubly inflated fat 
tiles are covered by three decagons, decagons at the positions 
A, D, E of Fig.~\ref{f_inf}(a). The decagon at the position~D 
can be one of the type-3,4,7,8,-9 and the decagon at 
the position~E can be one of 
the type-5,6,7,8,-9~\cite{foot2}.
However, For the four-fold deflated tiles, the types of 
decagons are uniquely determined for both fat and 
skinny shapes as shown in Fig.~\ref{f_tile_deco}(e).
Therefore, any dQUC model can be interpreted as an RPT model 
with quadratically inflated tiles.

\section{Concluding Remarks}

	We have shown that a dQUC model is an RPT model with the 
quadratically inflated tiles. Since an RPT model is a dQUC model,
we can conclude that the set of all dQUC models is the same as the 
set of all RPT models. However, mathematical equivalence between 
two sets does not imply that they are physically equivalent in 
constructing atomic models for quasicrystals. Constructing dQUC models 
is much easier than constructing RPT models for real quasicrystals. 
The atomic clusters used as building blocks in dQUC models are only 
one kind and much smaller than the building blocks of RPT 
models as shown here.

	Although the main contents of the paper are mathematical 
in nature, they can be applied in calculating some important 
physical quantities of dQUC models. For example, computing 
the density and stoichiometry of a dQUC model is complicated
due to overlaps if we directly calculate them. Since an explicit 
way to convert a dQUC model to an RPT model is provided here, 
such quantities can be easily calculated using the equivalent 
RPT model which has no overlaps between basic building blocks. 

This work was supported by Korea Research Foundation 
Grant (KRF-2001-041-D00062).

%\bibliography{All}

\begin{thebibliography}{10}
\bibitem{foot2}
This result can be directly understood 
from the inflation rule given at table~1 also. The columns of 
the inflated decagon at D and at E have type-(3,4,7,8,9)
and type-(5,6,7,8,9) respectively. 
However, to get the relationships between the further deflated
tiles and the typed decagons the method described 
in the text is more straightforward.
\end{thebibliography}
%\bibliographystyle{prsty}

\end{document}